\documentclass[12pt]{article}
\usepackage{amssymb}
\usepackage{amsmath, amsthm}
\newcommand{\be}{\begin{equation}}
\newcommand{\ee}{\end{equation}}
\begin{document}
\def\theequation{\arabic{section}.\arabic{equation}}
\begin{titlepage}
\title{What is the fate of  a black hole embedded in an expanding 
universe?}
\author{Valerio Faraoni$^1$, Changjun Gao$^2$, Xuelei Chen$^2$, and 
You-Gen Shen$^3$\\\\
{\small \it $^1$Physics Department, Bishop's University}\\
{\small \it 2600 College St., Sherbrooke, Qu\'{e}bec, Canada 
J1M~1Z7}\\
{\small \it $^2$ The  National Astronomical Observatories}\\
{\small \it Chinese
Academy of  Sciences, Beijing, 100012, China}\\
{\small \it $^3$ Shanghai Astronomical Observatory, Chinese Academy
of Sciences, Shanghai 200030, China}\\
{\small \it   Joint
Institute for Galaxy and Cosmology of SHAO and
  USTC, Shanghai 200030, China}
}

\date{} \maketitle
\thispagestyle{empty}
\vspace*{1truecm}
\begin{abstract}
Within a large class of exact solutions of the Einstein equations 
describing a black hole embedded in a Friedmann universe it is shown 
that, under certain assumptions, only those with comoving 
Hawking-Hayward quasi-local mass are 
generic, in the sense that they are late-time 
attractors.
 \end{abstract}
\begin{center}
\end{center} 
\begin{center} Keywords: black holes in cosmological backgrounds, 
time-dependent horizons.
\end{center} 
\end{titlepage} 
\clearpage 
\setcounter{page}{2} 

\section{Introduction}
\setcounter{equation}{0}

There is currently much interest in dynamical horizons 
\cite{dynhorizons} and solutions 
of the Einstein equations  
modelling dynamical black holes \cite{recent}. An example of such 
situations is 
that of a black hole embedded in an expanding 
Friedmann-Lemaitre-Robertson-Walker (FLRW) 
universe.  While the effect of the cosmic expansion on the local  
dynamics of astrophysical black holes is completely negligible today, 
this 
may 
not have been the case for primordial black holes in the very early 
universe \cite{primordialBHs}.  Moreover, the Hawking temperature and 
thermodynamics of 
time-dependent horizons appear to be interesting subjects for 
semiclassical 
gravity \cite{recent, SaidaHaradaMaeda, DiCriscienzoetal07, 
myPRDthermo}. 

With these motivations in mind, we have found exact solutions  
describing black holes embedded in FLRW spaces \cite{FaraoniJacques, 
Gaoetal}. For simplicity, and to reproduce the observed 
universe, we will limit ourselves to consider spatially flat 
expanding FLRW 
universes as backgrounds. These can be used as realistic solutions for 
certain 
situations and as toy models for others. Here we are concerned with two 
classes of solutions found in \cite{FaraoniJacques} and discussed in 
\cite{Gaoetal}. These solutions generalize the McVittie metric and can 
be written, in isotropic coordinates $\left( t,r,\theta,\varphi \right)$ 
as
\begin{eqnarray}
ds^2&=&-\frac{\left[1-\frac{M(t)}{2a\left(t\right)r}\right]^2}{
\left[1+\frac{M(t)}{2a\left(t\right)r}\right]^2} \, dt^2
  +{a^2\left(t\right)}\left[1+\frac{M(t)}{2a\left(t\right)r}
  \right]^4
   \nonumber\\ &&\nonumber \\
  &&\cdot\left(dr^2+r^2d\Omega^2 \right) \;,  \label{1}
 \end{eqnarray} 
where $d\Omega^2$ denotes the line element on the
unit 2-sphere.  The McVittie metric \cite{McVittie} corresponds to 
$M=M_0=$const., while here $M(t)$ is an {\em a priori} arbitrary 
function of the cosmic time $t$ which is positive and continuous with 
its first derivative. The constancy of $M$ in the McVittie 
metric expresses 
the McVittie assumption that $G_0^1=0$, {\em i.e.}, that the component 
of the stress-energy tensor $T_0^1=0$, hence there is no radial 
energy 
flow onto or from the central object (no radial accretion or excretion) 
\cite{McVittie}. As 
shown in 
\cite{FaraoniJacques}, the metric coefficient $M(t)$ is the 
Hawking-Hayward quasi-local mass \cite{Hawking}, which should be 
regarded as the 
physical mass of the central black hole. In conjunction with the 
 fact that the size of the McVittie central object does not 
change during the expansion of the universe,\footnote{The mass of the
 central object can, in principle, 
change because of two processes: the expansion of the object, which 
then swallows cosmic fluid, and a 
radial flow onto the object from far away or from the object.} the 
McVittie no-accretion 
condition 
$M(t)=$const. simply enforces the constancy of this mass.

Over the years, it 
became clear that, with the exception of the Schwarzschild-de Sitter 
metric, the McVittie spacetime can not describe a central black hole (or 
a regular strong field object) because of a spacelike singularity at 
$r=M_0/2$, corresponding to a diverging pressure of the cosmic perfect 
fluid sourcing this metric \cite{sus:1985, no:1998, no:1999}. 

Exact solutions discovered later \cite{Thakurta, fo:1995, SultanaDyer, 
ml:2006}, such as 
the Sultana-Dyer solution 
\cite{SultanaDyer}, if free from  singularities (other than the 
central black hole singularity and the usual cosmological ones), suffer 
from 
negative energy densities and, in addition, are limited in the type of 
FLRW background in which they can be embedded ({\em e.g.}, only an 
$a\propto t^{2/3}$ scale factor for the Sultana-Dyer solution), and in 
the type of matter that 
can source them ({\em e.g.}, a mixture of two perfect fluids, one of 
which is a null dust, for the Sultana-Dyer metric). 

The solutions~(\ref{1}) presented in \cite{FaraoniJacques} have the 
advantages of 
being free of singularities (apart from the central black hole 
singularity and the usual Big Bang or Big Rip cosmological 
singularities), and that the fluid source is relatively simple: a single 
imperfect fluid with a radial heat flux, described by the stress-energy 
tensor
\be
T_{\mu\nu}=\left( P+\rho \right) u_{\mu} u_{\nu} +Pg_{\mu\nu}
+ q_{\mu} u_{\nu} + q_{\nu} u_{\mu} \;,
\ee
where $u^{\mu}=\left( |g_{00}|^{-1/2}, 0,0,0, \right)$ is the fluid 
four-velocity and $q^{\mu}=\left( 0,q,0,0,\right)$ is the radial heat 
current.

In \cite{FaraoniJacques}, emphasis was put on a class of solutions with 
$ M(t)=M_0a(t) $, {\em i.e.}, with comoving quasi-local mass. These 
solutions possess a conformal Killing horizon and, in this sense, 
resemble the Sultana-Dyer solution which is constructed by conformally 
transforming the Schwarzschild metric with the scale factor of  
a dust-dominated universe as conformal factor, but requires a two-fluid  
source \cite{SultanaDyer}. The conformally expanding  solutions of 
\cite{FaraoniJacques} were also used in \cite{Gaoetal} and 
\cite{myPRDthermo}. In \cite{Gaoetal}, they were studied with 
emphasis on the 
behaviour of the black hole apparent horizon in  universes dominated 
by phantom dark energy with equation of state $P<-\rho$. A second  class  
of 
solutions with arbitrary function $M(t)$ was also discussed in 
\cite{Gaoetal}. Here we show that  the solutions of this second, and 
apparently more 
general, class can be attracted at late times toward  
the  ``comoving mass'' solutions during the expansion of the universe. 
Therefore, future research can safely focus on this much simpler class 
of comoving solutions.

\section{Comoving quasi-local mass solutions as late-time attractors}
\setcounter{equation}{0}

Following the notations of \cite{Gaoetal}, we begin by switching to the 
areal radius 
$\tilde{r} \equiv r\left( 1+\frac{M(t)}{2a(t)r} \right)^2 $ and then 
to its 
comoving version $R\equiv a\tilde{r}$, in terms of which the 
metric~(\ref{1}) is turned into the Painlev\'{e}-Gullstrand form
\begin{eqnarray}
 ds^2 & = & -\left[ 1-\frac{2 M}{R} -\frac{ \left( HR +
\dot{m}a\sqrt{ \frac{ \tilde{r}}{r}} \right)^2 }{ 1-\frac{2M}{R}
} \right] dt^2 \nonumber \\
&&\nonumber \\
& + &
\frac{dR^2}{1-\frac{2M}{R}}+R^2d\Omega^2 
 -   \frac{2}{1-\frac{2M}{R}} \left( HR+\dot{m}a\sqrt{
\frac{\tilde{r}}{r}} \, \right) dt \, dR \;,
\end{eqnarray}
where $m(t)\equiv M(t)/a(t)$. 
Defining  $ A(t,R) \equiv 1-2M/R $ and $ C(t,R) 
\equiv
HR +  \dot{m}a\sqrt{ \frac{\tilde{r}}{r} } $ and introducing a new 
time
coordinate $ T $ defined by 
\begin{equation} 
dT=\frac{1}{F}
\left( dt +\frac{C}{A^2-C^2}\, dR \right) \;, 
\end{equation}
where $F\left( T(t, R),R \right)$ is an integrating factor
that makes $d T $ an exact differential and satisfies 
\be
\partial_R\left( \frac{1}{F} \right)=\partial_t\left[ 
\frac{C}{F(C^2-A^2)} \right] \;,
\ee
 one cancels 
the cross terms in $dR \,dT $ and casts 
the line element in the Nolan gauge
\begin{eqnarray}
&&  ds^2 =- \left( 1-\frac{2M}{R} \right)\left[ 1- \frac{ \left(
HR+\dot{m}a \sqrt{ \frac{\tilde{r}}{r} } \, \right)^2}{\left(
1-\frac{2M}{R} \right)^2 } \right] F^2 dT^2 \nonumber \\
&& \nonumber \\
&& +\left( 1-\frac{2M}{R} \right)^{-1} \left[ 1-
 \frac{ \left( HR+\dot{m}a \sqrt{ \frac{\tilde{r}}{r} } \,
\right)^2}{\left( 1-\frac{2M}{R} \right)^2 } \right]^{-1} dR^2
  +R^2 d\Omega^2 \;. 
\end{eqnarray}
The black hole apparent horizon is located at the smallest root of the 
equation
\begin{equation}\label{100} 
HR + m a \left(1+\frac{m}{2r} \right) 
\left( \frac{\dot{M}}{M}-H \right)= 1 - \frac{2M}{R}  \;. 
\end{equation}
Since $r=r(R)$, this is an implicit equation for the horizon radius. The 
expression $\left( \frac{\dot{M}}{M}-H \right)=\frac{\dot{m}}{m} $ 
describes 
the 
deviation of the rate of change of the quasi-local black hole mass from 
the Hubble rate, {\em i.e.}, it  measures the deviation from stationary 
accretion {\em with respect to the background}. This expression vanishes 
for comoving mass solutions of the first class, which have 
$M(t)=M_0a(t)$.

We are now going to show that comoving mass solutions are generic 
under certain assumptions and 
all 
other solutions of the type~(\ref{1}) approach them at late times. 
We assume 1)~that the universe expands, 2)~that $m(t) \geq 0$, and 
3)~that this function is continuous with its first derivative.
Let 
us use the radial variable $\tilde{r}\equiv R/a$. Then, 
eq.~(\ref{100}) 
becomes 
\be\label{200}
H \tilde{r}+ \frac{2m}{\tilde{r} a} = -\dot{m}\left( 1+\frac{m}{2r} 
\right) + \frac{1}{a} 
\;.
\ee
Since $m\geq 0$, the left hand side is clearly non-negative at all 
times, so $\dot{m} 
\left( 1+\frac{m}{2r}\right)<\frac{1}{a}$. Therefore, in an expanding 
universe in which $ a\rightarrow + \infty$, and given that 
$1+\frac{m}{2r}>0$, it is $ \dot{m}_{\infty} \equiv \lim_{t\rightarrow 
+\infty} \dot{m}(t) \leq 0$. If $\dot{m}_{\infty}=0$, the black hole 
becomes asymptotically comoving. Let us focus on the case 
$\dot{m}_{\infty}<0$. Then, there exists a time $\bar{t}$ such that, for 
all times $t>\bar{t}$, it is $\dot{m}(t)<0$. There are only two 
possibilities in this case: since $ m(t) \geq 0$, 
either\\
a) $m(t)$ reaches the value zero at a finite time $t_*$ with derivative 
$\dot{m}_*\equiv \dot{m}(t_*) < 0$, or\\
b) $m(t) \rightarrow m_0=$const. with $\dot{m}(t)\rightarrow 0$, {\em 
i.e.}, $m(t)$ has a horizontal asymptote.

In case~a) one has, at $t=t_*$, ~$ HR =\left| \dot{m}_*\right| a +1 $, 
which yields the radius of the black hole apparent horizon at $t_*$
\be
r_* \equiv r_{horizon}(t_*)= \frac{1}{H(t_*)} 
\left(|\dot{m}_{*}| +\frac{1}{a} \right) \;.
\ee
Late in the history of the universe, we have  a black hole of zero 
quasi-local mass $ M(t_*)=a(t_*)m(t_*)$ but finite radius $r_*$. As time 
evolution  continues, one would have negative mass $M$ and 
finite radius of the black hole apparent horizon. This does not make 
sense physically and, therefore, the case 
$m(t_*)=0$ with $m(t>t_*)<0$ is ruled out.

We are left with  case~b) in which $\dot{m}(t) \rightarrow 0$ at late 
times ({\em i.e.}, $t\rightarrow +\infty$ if the cosmic expansion 
continues forever, or $ t\rightarrow t_{rip} $ if a Big Rip occurs at 
the time $t_{rip}$). The physical meaning is that, at late times, the 
rate 
of increase of the black hole mass is at most the Hubble rate and the 
black hole becomes comoving. This conclusion is, of course, not valid at 
early 
times, at which the term $1/a$ in eq.~(\ref{200}) does not tend to zero.

As a special case of~b), it is possible that $m_0$ is zero, in which 
case the solution reduces to a FLRW universe without 
inhomogeneities and 
can be interpreted as a black hole that evaporates 
completely\footnote{An obstacle to this interpretation is that 
the radial flow considered in these solutions is not described by  
a null vector. A generalization will be pursued elsewhere.}. This 
possibility is non-trivial from the physical point of view.

Physically relevant situations in which the black hole does not become 
comoving, which are not included in the previous description, may arise 
if the assumptions are relaxed. For example, if the 
assumption of continuity of $\dot{m}(t)$ is dropped, one can contemplate 
the situation in which $m(t) \rightarrow 0$ in a finite time $t_{ev}$ 
and 
\be
\dot{m}(t) \left\{ 
\begin{array}{cc} 
< 0 \;\;\;\;\; \mbox{if}\; t\leq t_{ev} \;, \\
=0 \;\;\;\;\; \mbox{if}\; t> t_{ev} \;,
\end{array}
\right.
\ee
Such a spacetime would have a continuous metric, discontinuous 
Christoffel symbols, and distributional curvature (in analogy to  
exact $pp$-waves) and could represent a black hole evaporating as 
$m\rightarrow 0 $ when $t\rightarrow t_{ev}$. A detailed study of this 
possibility will be pursued elsewhere.

\section{Discussion and conclusions}
\setcounter{equation}{0}

The two classes of exact solutions of the Einstein equations recently 
proposed in \cite{FaraoniJacques,Gaoetal} and  describing a black 
hole embedded in a spatially flat FLRW universe and accreting cosmic 
fluid, are of interest to 
study dynamical horizons and their thermodynamics. Such solutions are 
useful as  testbeds for various conjectures on time-dependent horizons, 
and are relatively rare. It is therefore important to look for simple 
solutions which do not suffer from problems such as unphysical 
singularities, negative energy densities, or being sourced by exotic  
forms of matter that could hide the physics under investigation.

Under the assumptions 1)~that the universe expands; 2)~that the mass 
parameter $m$ is non-negative; and 3)~ that the function $m(t)$ is 
continuous with its first derivative, we have 
shown  
that only the first class of solutions considered in 
\cite{FaraoniJacques,Gaoetal}, in which the Hawking-Hayward quasi-local 
mass 
is comoving, $M(t)=m_0 a(t)$, is generic, in the sense that these 
solutions act 
as late-time attractors for all the 
solutions of the second class (exceptions are black holes with mass 
asymptotically going to zero, which cannot be called ``comoving''). 
Therefore, future research can focus on 
the first class of solutions, which are simpler (that they are much  
simpler than solutions of the second class was demonstrated in the study 
of their black hole and cosmic  apparent horizons in \cite{Gaoetal}).

Apart from the interest in time-dependent horizons and their 
thermodynamics, a lesson to be learned is that, in an expanding 
universe, self-gravitating objects eventually tend to participate 
in the global expansion and to align their 
evolutionary dynamics with that of the cosmic substratum. The situation 
studied here for black holes is very similar to that investigated for   
wormholes in Ref.~\cite{FI}. There, using exact solutions describing a 
wormhole embedded in a FLRW universe, it was found 
that  even if  a wormhole starts expanding much faster (or much slower) 
than the cosmic substratum, eventually it catches up with the cosmic 
expansion and becomes comoving.

At present, it is not clear whether the metric~(\ref{100}) is the most 
general spherically symmetric solution describing  a black hole embedded 
in a spatially flat FLRW background, in the same sense that the 
Schwarzschild solution is the most general vacuum, spherically 
symmetric and asymptotically flat black hole metric.  This can only be 
decided by  a separate analysis and verified by perturbation studies, 
which will be pursued elsewhere.

\section*{Acknowledgments}

We acknowledge a referee for insightful remarks. 
This work is supported by the Natural Sciences and Engineering
Research Council of Canada, the National Science
Foundation of China under the Distinguished Young Scholar Grant
10525314, the Key Project Grant 10533010, and Grants No. 10575004,
10573027 and 10663001; by the Shanghai Natural Science Foundation
under Grant No. 05ZR14138; by the Chinese Academy of Sciences under
grant KJCX3-SYW-N2; and by the Ministry of Science and Technology
under the National Basic Sciences Program (973) under grant
2007CB815401.

\vskip1truecm

\newcommand\AL[3]{~Astron. Lett.{\bf ~#1}, #2~ (#3)}
\newcommand\AP[3]{~Astropart. Phys.{\bf ~#1}, #2~ (#3)}
\newcommand\AJ[3]{~Astron. J.{\bf ~#1}, #2~(#3)}
\newcommand\APJ[3]{~Astrophys. J.{\bf ~#1}, #2~ (#3)}
\newcommand\APJL[3]{~Astrophys. J. Lett. {\bf ~#1}, L#2~(#3)}
\newcommand\APJS[3]{~Astrophys. J. Suppl. Ser.{\bf ~#1},
#2~(#3)} \newcommand\JCAP[3]{~JCAP. {\bf ~#1}, #2~ (#3)}
\newcommand\LRR[3]{~Living Rev. Relativity. {\bf ~#1}, #2~ (#3)}
\newcommand\MNRAS[3]{~Mon. Not. R. Astron. Soc.{\bf ~#1},
#2~(#3)} \newcommand\MNRASL[3]{~Mon. Not. R. Astron. Soc.{\bf
~#1}, L#2~(#3)} \newcommand\NPB[3]{~Nucl. Phys. B{\bf ~#1},
#2~(#3)} \newcommand\PLB[3]{~Phys. Lett. B{\bf ~#1}, #2~(#3)}
\newcommand\PRL[3]{~Phys. Rev. Lett.{\bf ~#1}, #2~(#3)}
\newcommand\PR[3]{~Phys. Rep.{\bf ~#1}, #2~(#3)}
\newcommand\PRD[3]{~Phys. Rev. D{\bf ~#1}, #2~(#3)}
\newcommand\RMP[3]{~Rev. Mod. Phys.{\bf ~#1}, #2~(#3)}
\newcommand\SJNP[3]{~Sov. J. Nucl. Phys.{\bf ~#1}, #2~(#3)}
\newcommand\ZPC[3]{~Z. Phys. C{\bf ~#1}, #2~(#3)}


\begin{thebibliography}{99}

\bibitem{dynhorizons}
S.A. Hayward, S. Mukohyama, and M.C.
Ashworth, {\em Phys. Lett. A} {\bf 256}, 347 (1999); S.A. Hayward, {\em 
Class. Quantum Grav.} {\bf 15}, 3147 (1998);  A. Ashtekar and B. 
Krishnan, {\em Phys.  Rev. Lett.} {\bf 89}, 261101 (2002).

\bibitem{recent} 
X. Zhang, arXiv: 0708.1408; D. Brill, G. Horowitz, D. Kastor, J. Traschen,
{\em Phys. Rev. D} {\bf 49}, 840 (1994);  A. Ashtekar and G.J. Galloway, 
arXiv:grqc/0503109;
A. Ashtekar and B. Krishnan, {\em Living Rev. Rel.} {\bf  7}, 10 (2004); 
{\em Phys.
Rev. D} {\bf 68}, 104030 (2003); {\em Phys. Rev. Lett.} {\bf 89}, 261101 
(2006); A. Ashtekar and A. Corichi, {\em Class. Quantum Grav.} {\bf 17},  
1317  (2000);
A. Ashtekar, A. Corichi, and D. Sudarsky, {\em Class. Quantum Grav.} 
{\bf 20}, 3413 (2003); A. Ashtekar, J. Engle, T. Pawlowski and C. Van 
Den Broeck, {\em Class. Quantum Grav.} {\bf 21}, 2549 (2004); A. 
Ashtekar, arXiv:gr-qc/0306115; A. Ashtekar, C. Beetle, and J. 
Lewandowski, {\em Class. Quantum Grav.} {\bf 19}, 1195 (2002);  {\em 
Phys. Rev. D} {\bf  64},  044016
(2001); A. Ashtekar, C. Beetle, O. Dreyer, S. Fairhurst, B.
Krishnan, J. Lewandowski, and J. Wisniewski, {\em Phys. Rev. Lett.} {\bf  
85}, 3564 (2000); A. Ashtekar, S. Fairhurst and B. Krishnan, {\em Phys. 
Rev.  D} {\bf 62}, 104025 (2000); A. Ashtekar, C. Beetle and S. 
Fairhurst, {\em Class. Quantum Grav.} {\bf  17}, 253 (2000); {\bf 16}, 
L1 (1999); A.B. Nielsen  and M. Visser, {\em Class.  Quantum Grav.} {\bf  
23}, 4637 (2006); 
D. Kotawala,  S. Sarkar, and T. Padmanabhan, {\em Phys. Lett. B} {\bf  
652}, 338 (2007);
 M. Nadalini, L. Vanzo, and S. Zerbini, arXiv:07010.2474;  J.A. de 
Freitas Pacheco and J.E. Horvath, arXiv:0709.1240;  D.C. Guariento, J.E. 
Horvath, P.S. Custodio, and J.A. de Freitas Pacheco, arXiv:0711.3641; 
P.S. Custodio and J.E.  Horvath, {\em Int. J. Mod. Phys. D} {\bf  14}, 
257 (2005); 
E. Babichev, V. Dokuchaev,  and Yu.  Eroshenko,   {\em Phys. Rev. Lett.}  
{\bf 93}, 021102 (2004); 
I. Booth, {\em Can. J. Phys.} {\bf  83}, 1073 (2005);
G. Izquierdo and D. Pavon, {\em Phys. Lett. B} {\bf 639}, 1 (2006);  S.
Chen and J. Jing, {\em Class. Quantum Grav.} {\bf 22}, 4651 (2005); T.
Clifton, D.F. Mota, and J.D. Barrow, {\em Mon. Not. R. Astron. Soc.}
{\bf 358}, 601 (2005);
N. Sakai and J.D. Barrow, {\em Class. Quantum Grav.} {\bf 18}, 4717 
(2001); 
S.A. Hayward, {\em Phys. Rev. D} {\bf 70}, 
104027 (2004);  {\em Phys. Rev. Lett.} {\bf 93}, 251101 (2004);
arXiv:gr-qc/0008071;
{\em Phys. Rev. Lett.} {\bf 81}, 4557 (1998); {\em Phys. Rev. D} {\bf 
53}, 1938 (1996);
{\em Class. Quantum Grav.} {\bf 11}, 3025 (1994); {\em Phys. Rev. D} 
{\bf  49}, 6467
(1994);
 arXiv:gr-qc/9303006; S. Mukohyama and S.A. Hayward, {\em Class.
Quantum Grav.} {\bf 17}, 2153 (2000); 
H. Maeda, arXiv:0704.2731; G. Kang, {\em Phys. Rev. D} {\bf 54},
7483 (1996); T. Jacobson and G. Kang, {\em Class. Quantum Grav.} {\bf 
10}, L201 (1993); 
C.C. Dyer and E. Honig, {\em J. Math. Phys.} {\bf 20}, 409 
(1979);
S. Nojiri and S.D.  Odintsov, {\em Phys. Rev. D} {\bf 70}, 103522 
(2004).

\bibitem{primordialBHs} Y.B. Zel'dovich and I.D. Novikov, {\em 
Sov. Astr. A.J.} {\bf 10}, 602 (1967); 
S.W. Hawking, {\em Phys.  Rev. Lett.} {\bf 26}, 1344 (1971); 
{\em Mon.  Not. R. Astr. Soc.} {\bf 152}, 75 (1971);
B.J. Carr and S.W.  Hawking, {\em Mon. Not. R. Astr. Soc.} {\bf 
168}, 399 (1974); 
N. Sakai and J.D. Barrow, {\em Class. Quant. Grav.} {\bf 18}, 4717 
(2001).

\bibitem{SaidaHaradaMaeda} H. Saida, T. Harada, and H. Maeda,
{\em Class. Quantum Grav.} {\bf 24}, 4711 (2007);  
M. Nozawa and H. Maeda, arXiv:0710.2709.

\bibitem{DiCriscienzoetal07} R. Di Criscienzo, M. Nadalini, L.
Vanzo, S. Zerbini, and G. Zoccatelli, {\em Phys. Lett. B} {\bf  657}, 
107 (2007).

\bibitem{myPRDthermo} V. Faraoni, {\em Phys. Rev. D} {\bf  76}, 104042 
(2007).

\bibitem{FaraoniJacques} V. Faraoni and A. Jacques, {\em Phys. Rev. D} 
{\bf 76}, 063510 (2007).

\bibitem{Gaoetal}  C. Gao, X. Chen, V. Faraoni and Y. G. Shen, {\em 
Phys. Rev. D} {\bf 78}, 024008 (2008).

\bibitem{McVittie} G.C. McVittie, {\em Mon. Not. R. Astron. Soc.} {\bf  
93}, 325 (1933).

\bibitem{Hawking} S.W. Hawking, {\em J. Math. Phys.} {\bf 9}, 
598 (1968); S.A. Hayward, {\em Phys. Rev. D} {\bf 49},  831 (1994).

\bibitem{sus:1985} R. Sussman, {\em Gen. Rel. Grav.} {\bf 17}, 251 
(1985);
M. Ferraris, M. Francaviglia, and A. Spallicci, {\em  Nuovo Cimento}
{\bf 111B}, 1031 (1996); 
B.C. Nolan, {\em Class. Quantum Grav.} {\bf 16}, 1227 (1999).

\bibitem{no:1998} B.C. Nolan, {\em Phys. Rev. D} {\bf 58}, 064006 
(1998).

\bibitem{no:1999} B.C. Nolan, {\em Class. Quantum Grav.} {\bf 16}, 3183
(1999).

\bibitem{Thakurta} S.N.G. Thakurta, {\em Indian J. Phys.} {\bf 55B}, 
304 (1981).

\bibitem{fo:1995} O.A. Fonarev, {\em Class. Quantum Grav.} {\bf 12}, 
1739 (1995).

\bibitem{SultanaDyer} J. Sultana and C.C. Dyer, {\em Gen. Rel. Grav.}
{\bf 37}, 1349 (2005).

\bibitem{ml:2006} M.L. McClure and C.C. Dyer, {\em Class. Quantum
Grav.} {\bf 23}, 1971 (2006); {\em Gen. Rel. Gravit.} {\bf 38}, 1347 
(2006).

\bibitem{FI} V. Faraoni and W. Israel, {\em Phys. Rev. D} {\bf 71}, 
064017  (2005).


\end{thebibliography}
\end{document}